%% file: main.tex
\documentclass[format=acmsmall,screen=true,nonacm]{acmart}

\usepackage{booktabs}
\usepackage[ruled]{algorithm2e}
\graphicspath{{Figures}}
\DeclareGraphicsExtensions{.pdf,.png}
\usepackage{bigstrut}
\usepackage{multirow}
\usepackage{multicol}

\newcommand{\p}[1]{\smallskip \noindent \textbf{{#1}.}}
\newcommand{\eq}[1]{Equation~(\ref{eq:#1})}
\newcommand{\fig}[1]{Figure~\ref{fig:#1}}

\usepackage{graphicx}
\graphicspath{{Figures}}
\DeclareGraphicsExtensions{.pdf,.png}

\usepackage{csquotes}
\usepackage{balance}
\usepackage{amsmath}

\setcopyright{none}

\begin{document}

\title[Here's What I've Learned: Asking Questions that Reveal Reward Learning]{\textit{Here's What I've Learned:} \\ Asking Questions that Reveal Reward Learning}
%Learning Robot Trajectories from Physical Human Corrections for One-Shot Tasks
%LQR + Learning for Real-Time Optimal Responses to Physical Human Interactions
%Leveraging Physical Human Interactions to Correct Robot Trajectories in Real-Time

\author{Soheil Habibian}
\orcid{0000-0003-4103-4664}
\email{habibian@vt.edu}

\author{Ananth Jonnavittula}
\orcid{0000-0002-0711-2051}
\email{ananth@vt.edu}

\author{Dylan P. Losey}
\orcid{0000-0002-8787-5293}
\email{losey@vt.edu}

\affiliation{
  \institution{Virginia Tech}
  \department{Department of Mechanical Engineering}
  \streetaddress{635 Prices Fork Rd}
  \city{Blacksburg}
  \state{VA}
  \postcode{24060}
  \country{USA}
  }

\renewcommand\shortauthors{S. Habibian et al.}

\begin{abstract}
Robots can learn from humans by asking questions. In these questions the robot demonstrates a few different behaviors and asks the human for their favorite. But how should robots choose which questions to ask? Today’s robots optimize for \textit{informative} questions that actively probe the human’s preferences as efficiently as possible. But while informative questions make sense from the robot’s perspective, human onlookers often find them arbitrary and \textit{misleading}. For example, consider an assistive robot learning to put away the dishes. Based on your answers to previous questions this robot knows where it should stack each dish; however, the robot is unsure about right height to carry these dishes. A robot optimizing only for informative questions focuses purely on this height: it shows trajectories that carry the plates near or far from the table, regardless of whether or not they stack the dishes correctly. As a result, when we see this question, we mistakenly think that the robot is still confused about where to stack the dishes! In this paper we formalize active preference-based learning from the human’s perspective. We hypothesize that --- from the human’s point-of-view --- the robot’s questions \textit{reveal} what the robot has and has not learned. Our insight enables robots to use questions to make their learning process \textit{transparent} to the human operator. We develop and test a model that robots can leverage to relate the questions they ask to the information these questions reveal. We then introduce a trade-off between informative and revealing questions that considers both human and robot perspectives: a robot that optimizes for this trade-off actively gathers information from the human while simultaneously keeping the human up to date with what it has learned. We evaluate our approach across simulations, online surveys, and in-person user studies. We find that robots which consider the human's point of view learn just as quickly as state-of-the-art baselines while also communicating what they have learned to the human operator. Videos of our user studies and results are available here: \url{https://youtu.be/tC6y_jHN7Vw}.
\end{abstract}


%
% End generated code
%


\maketitle

\input{intro}
\input{related}

\input{problem}
\input{model}
\input{amt}
\input{questiontypes}
\input{simulations}
\input{userstudy}
\input{conclusion}
\input{limitations}

\balance
\bibliographystyle{ACM-Reference-Format}
\bibliography{sample-bibliography}

\end{document}

%% file: intro.tex
\section{Introduction}

\begin{figure*}[t]
	\begin{center}
		\includegraphics[width=1\columnwidth]{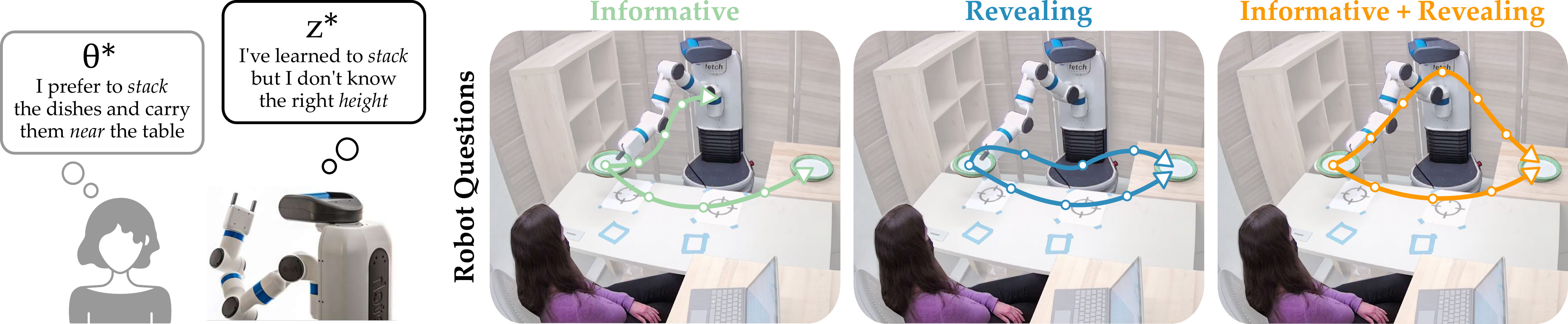}
		\caption{A robot asking multiple-choice questions to learn the human's reward function. The human's true reward is parameterized by $\theta^*$, while the robot's current understanding of that reward is captured by $z^*$. Within the state-of-the-art, the robot optimizes for \textcolor[rgb]{0.63, 0.83, 0.64}{\textbf{Informative}} questions that maximize its information gained about $\theta^*$. We formulate the opposite perspective, in which the robot purposely chooses \textcolor[rgb]{0.15, 0.55, 0.74}{\textbf{Revealing}} questions to communicate its learning $z^*$ to the human. We then combine both objectives: \textcolor[rgb]{1.0, 0.6, 0.0}{\textbf{Informative + Revealing}} robots trade-off between robot and human perspectives, and select questions that both elicit user information and reveal the robot's learning. Within this example the robot's question should emphasize that it already knows to stack the dishes while gathering clear feedback about how high to carry these dishes.}
		\label{fig:front}
	\end{center}

\end{figure*}

Imagine that you are teaching a robot arm to put away the dishes (see \fig{front}). You want the robot to carry these dishes close to the table --- so they will not break if they fall --- and to stack the dishes on top of each other. A natural way for the robot to learn your preferences is by \textit{asking questions}. In each question the robot shows you two different ways to put away the dishes and asks for your favorite. Based on your answers so far, the robot is (a) confident where it should stack the dishes and (b) unsure how high above the table to carry these dishes. Of course, the robot knows what it has learned --- but how do \textit{you know} whether or not this robot understands your preferences?

Recent work on active preference-based learning explores how robots can leverage questions to learn from humans \cite{sadigh2017active, biyik2019asking, biyik2020learning, ibarz2018reward, christiano2017deep, brown2019extrapolating, daniel2014active}. Within the state-of-the-art the robot actively selects questions to gather as much as information as possible. Although this makes sense from the robot's point of view, it often results in questions that the human finds misleading. In our running example shown in \fig{front}, the most \textit{informative} question is for the robot to show one trajectory that moves close to the table (and then stacks the dishes), and a second trajectory that always remains high above the table (and does not stack the dishes as a result). This question is informative because it elicits clear feedback about the preferred height, which the robot is currently unsure about. But the question also \textit{misleads} the human: when we see this question, we incorrectly think the robot is still unsure about both the preferred height and whether to stack the dishes.

In this paper we consider how robot questions influence the human's perception of the robot. Our key insight is that --- from the human's perspective --- 
\begin{center}
\textit{The questions that the robot asks} reveal \textit{what the robot does and does not know.}
\end{center}
Applying this insight enables robots to actively pick questions that keep the human up-to-date with the robot's learning. Put another way, now the robot can purposely select questions to reveal its learning process. Returning to our running example in \fig{front}, the most \textit{revealing} question is for the robot to show two similar trajectories that both accurately stack the dishes while carrying those dishes at varying heights. This question is revealing because it expresses what the robot knows (stacking the dishes) and what the robot is unsure about (the carrying height). But while this question no longer misleads the human, the human's answer to this question is not very useful for the robot. Looking again at \fig{front}, no matter which trajectory we choose, our answer tells the robot \textit{no new information} about whether to carry the dishes high or low. 

We therefore introduce a spectrum between the robot's perspective (optimizing for informative questions) and the human's perspective (optimizing for revealing questions). Taking both human and robot perspectives into account leads to \textit{informative and revealing} questions. Returning to our example, an informative and revealing question consists of one trajectory that always remains close to the table and stacks the dishes, and a second trajectory that initially carries the dishes high in the air before returning to the table to stack the dishes. This question gathers meaningful information about the human's preferred height (i.e., the two options have different heights), but simultaneously reveals what the robot currently knows (i.e., both options correctly stack the dishes). Hence, the benefit of our proposed approach is that it identifies questions which efficiently gain information while also communicating what the robot is learning back to the user. As we will show, this lets the robot learn just as efficiently as state-of-the-art active learning baselines, but now with the added dimension of proactively bringing the human into the learning loop.

Communicating what the robot is learning is an important step towards interpretable human-robot interaction \cite{arrieta2020explainable}. As the human infers what their robot partner has learned (and what it is still confused about) the human can more confidently predict how the robot will behave. More generally, we view the questions that the robot asks as one avenue towards establishing \textit{trust} and making robot learning \textit{transparent} to human operators. Overall, we make the following contributions:

\p{Formalizing Revealing Questions} We formulate how humans learn from robot questions, and introduce a cognitive model that robots can leverage to relate the questions they ask to the information those questions reveal. We then conduct an online survey to test our model and verify that revealing questions successfully convey robot learning to the human.

\p{Combining Informative and Revealing Questions} We derive an optimization approach for selecting informative and revealing questions that gather human preferences while conveying robot learning. This proposed approach considers both human and robot perspectives, and allows designers to trade-off along a spectrum between both extremes. We theoretically prove that this algorithm converges towards asking revealing questions over repeated interactions.

\p{Comparing to Baselines} We conduct a series of simulations and in-person user studies to compare our approach to state-of-the-art alternatives. We find that when robots only optimize for informative questions, these questions tell the \textit{human} no more about the robot's learning than simply asking random questions. Conversely, when robots only optimize for revealing questions, these questions tell the \textit{robot} no more about the human's preferences than random questions. But when the robot leverages our informative and revealing approach it satisfies both objectives: the robot learns efficiently while also communicating what it has learned.

%% file: related.tex
\section{Related Work}

Robots learn what their human partner wants by observing the human's inputs. These inputs can take many different forms \cite{jeon2020reward}, from physical demonstrations to verbal commands to pairwise comparisons. We focus on the latter, and explore how robots should ask questions to learn the human's underlying reward function. We consider the \textit{bi-directional} flow of information: the robot learning from the human and the human inferring what their robot is learning. Our research builds on prior work that studies how robots actively select questions, as well as research on how robots communicate what they have learned to nearby human operators. 

\subsection{Learning Reward Functions from Questions}

\p{Inverse Reinforcement Learning} We consider scenarios where the human has in mind a reward function and the robot is trying to learn this reward function from the human's behavior. One common approach here is for the human to input expert \textit{demonstrations} --- i.e., for the human to kinesthetically show the robot how it should perform the task. Inverse reinforcement learning enables robots to infer the human's reward function from these demonstrations \cite{abbeel2004apprenticeship, ziebart2008maximum, ramachandran2007bayesian, osa2018algorithmic}. But often it is challenging for humans to \textit{accurately} demonstrate what they want. Physical limitations constrain the human \cite{jonnavittula2020know, tucker2020preference}, particularly when guiding the robot requires carefully coordinating the motion of multiple interconnected joints \cite{akgun2012keyframe, bajcsy2018learning}. Cognitive biases also affect the quality of human demonstrations: for instance, people prefer for autonomous cars to be more risk-averse than their own driving demonstrations \cite{basu2017you, kwon2020humans}. So while we draw from previous work on inverse reinforcement learning, we now apply this theory to a different form of human input --- the human's answers to the robot's questions.

\p{Active Preference-Based Learning} Each question consists of the robot showing the human a few trajectories and asking the human to compare these options. For example, the robot could display three or more trajectories, and ask the human to rank these trajectories in order of personal preference \cite{brown2019extrapolating, jain2015learning}. Similar to prior research, within our simulations and user studies we specifically consider \textit{pairwise} comparisons where the robot only shows \textit{two} options \cite{sadigh2017active, jeon2020reward, biyik2019asking, biyik2020learning, wilde2020active, ibarz2018reward, tucker2020preference, christiano2017deep}. Of course, there are an infinite number of possible questions; how does the robot choose which of these questions to ask? Existing work on active preference-based learning enables the robot to intelligently select questions that efficiently learn about the human's reward \cite{racca2018active,ailon2012active,daniel2014active,shah2020interactive,thomason2017opportunistic}. One state-of-the-art approach here is for the robot to select questions that optimize for the expected information gain \cite{biyik2019asking, biyik2020learning}, so that no matter which answer the human chooses, the robot gathers meaningful feedback about the human's reward function. Robots proactively select these questions to accelerate their learning and reduce the number of queries the human must answer.

\p{User-Friendly Questions} Unfortunately, the questions generated by a robot trying to optimize its learning are often misleading or confusing to the human. Most related to our approach is research on how robots that actively choose questions should account for the \textit{human-in-the-loop}. Recent works recognize that human answers are inevitably noisy and imperfect \cite{holladay2016active}, and so we can improve both the robot's learning and human's experience by purposely asking easier questions \cite{biyik2019asking, biyik2020learning, cakmak2012designing, racca2019teacher}, by allowing users to explain their answers \cite{basu2018learning}, or by explicitly accounting for the time constraints that humans face when answering questions \cite{bullard2019active}. Each of these works takes a step towards user-friendly interactions. But while today's robots account for how humans answer questions, they still only consider \textit{one direction of information transfer}: from the human to the robot. Unlike prior research, we now focus on bi-direction information transfer. We hypothesize that the questions that the robots asks influence what the human thinks the robot knows --- enabling us to \textit{close the loop} and leverage questions to convey robot learning back to the human.

\subsection{Communicating Robot Learning through Robot Motions} 

\p{Legible Motion} Our key insight is that humans make inferences about the robot's latent state based on the robot's questions. Recalling that these questions are composed of a set of trajectories, our insight claims that humans infer what the robot is thinking based on the motions the robot is showing. Prior work from cognitive science suggests that humans naturally ascribe goals, plans, and objectives to robot motion \cite{schmidt1978plan, cohen1981beyond, baker2009action, hellstrom2018understandable}. For instance, if we see a robot arm moving directly towards a cup, we may infer that the robot wants to reach that cup. Recent research within robotics leverages this insight to purposely choose motions that communicate the robot's reward function \cite{huang2019enabling, losey2019robots, dragan2013legibility} or the robot's capabilities \cite{tellex2014asking, kwon2018expressing}. Similarly, our approach takes advantage of the robot's behavior (i.e., the different trajectories in each question) to communicate what the robot has learned (i.e., the robot's estimate of the human's reward function).

\p{Multimodal Feedback} We recognize that the robot's physical behavior is one of many modalities for conveying feedback to the human. Robots can also leverage separate channels such as augmented reality \cite{rosen2019communicating, walker2018communicating} or natural language \cite{tellex2014asking, nikolaidis2018planning} to relay information. But these different feedback modalities are not mutually exclusive; instead, they form complementary ways to intuitively bring the human into the learning loop. Although here we will specifically study how the robot's questions (i.e., the robot's physical behavior) communicates information, we ultimately see this feedback as part of the larger spectrum of interpretable robot learning \cite{arrieta2020explainable}.

%% file: problem.tex
\section{Formalizing Questions from Robot and Human Perspectives} \label{sec:problem}

\begin{table}
    \caption{Key Variables and their Definition}\vspace{-1em}
    \begin{tabular}{ r l }
        \hline 
        $\xi \in \Xi$ & robot trajectory \bigstrut[t] \\
        $Q = \{\xi_1, \ldots, \xi_N\}$ & question where the robot shows $N$ trajectories $\xi_1, \ldots, \xi_N$ \\
        $q \in Q$ & human's answer to the robot's question\\
        $\theta^* \in \mathbb{R}^d$ & human's preferences the robot is trying to learn\\
        $b_{\mathcal{R}}(\theta)$ & robot's learned belief over what $\theta^*$ is \\
        $z^* \in Z$ & parameterization of the robot's belief $b_{\mathcal{R}}$ \\
        $b_{\mathcal{H}}(z)$ & humans's belief over what $z^*$ is, i.e., over what the robot has learned \bigstrut[b] \\
        \hline
        \label{table:def}
    \end{tabular}
\end{table}

In this section we formulate our problem setting from \fig{front}. Our goal is for a robot partner to select questions that proactively gather information from the human while also revealing what the robot is learning. We start with the \textit{robot's perspective}: this robot is asking questions to learn the user's preferred behavior. We then introduce the \textit{human's perspective}: this human reasons about what the robot has learned based on the questions the robot asks. We list the key variables we use to characterize our bi-directional information exchange in Table~\ref{table:def}.

\p{Robot Perspective} The robot is learning how it should behave. More formally, the robot is trying to learn the human's \textit{reward function} $R$. This reward function assigns scores to \textit{trajectories} $\xi \in \Xi$, so that learning $R$ enables the robot to identify the optimal, highest scoring trajectory to deploy.

Following prior work on inverse reinforcement learning \cite{abbeel2004apprenticeship, ziebart2008maximum, osa2018algorithmic}, we formulate the human's reward function as a linear combination of features $f$ weighted by preferences $\theta$:
\begin{equation} \label{eq:P1}
    R(\xi, \theta) = \theta \cdot f(\xi), \quad \| \theta \| = 1
\end{equation}
Here the \textit{features} $f: \Xi \rightarrow \mathbb{R}^d$ are task-related metrics that the robot measures. For instance, in our motivating example one feature is the height of the robot's end-effector. Although the robot knows these features, it does not know which features the human prefers. \textit{Preferences} are captured by the unit vector $\theta \in \mathbb{R}^d$, and we denote the human's true preferences as $\theta^*$. Returning to our motivating example, the human prefers for the robot to move close to the table. Put in terms of \eq{P1}, this translates to $\theta_{height}^* < 0$, which indicates that higher trajectories (i.e., trajectories with an increased height feature) will receive less reward from the human.

\p{Questions} When the robot first interacts with a user it does not know their actual preferences $\theta^*$. But robots can intuitively learn the user's preferences by asking questions. These questions are multiple choice: the robot shows the human $N$ different trajectories, and asks the human to pick one trajectory that best matches their preferred behavior. Here $Q = \{\xi_1, \ldots, \xi_N\}$ is the robot's \textit{question} and $q \in Q$ is the human's \textit{answer}. The robot reasons across the questions it has asked and the answers the human has given to infer a continuous probability distribution over what $\theta^*$ is. We refer to this distribution as the \textit{robot's belief}. After $i$ questions, the robot's belief is:
\begin{equation} \label{eq:P2}
    b_{\mathcal{R}}^{i+1}(\theta) = P(\theta \mid Q_1, \ldots, Q_i, q_1, \ldots, q_i)
\end{equation}
By definition, $b_{\mathcal{R}}^{i+1}(\theta)$ expresses how likely it is that the user's true preferences are $\theta$ (i.e., $\theta^*=\theta)$ given the robot asked questions $Q_1, \ldots, Q_i$ and the human chose answers $q_1, \ldots, q_i$.

\p{Human Perspective} Prior work studies Equations~(\ref{eq:P1}) and (\ref{eq:P2}), and shows how the \textit{robot} should actively select questions to infer $\theta^*$ \cite{sadigh2017active, biyik2019asking, basu2018learning, ailon2012active, holladay2016active}. Moving beyond prior work, we now consider how these questions affect the \textit{human's perception} of the robot's learning. When the robot asks questions about a specific feature of the task (i.e., each $\xi \in Q$ carries the dishes at a different height), the human thinks the robot is uncertain about this feature. Conversely, when the robot displays questions that keep specific features consistent (i.e., the robot stacks the dishes in every $\xi \in Q$), the human may guess the robot is confident about this feature.

We hypothesize that the questions the robot asks reveal what the robot has learned. Recall that the robot is learning a belief $b_{\mathcal{R}}$. If the human could directly observe this belief, the human would understand exactly what the robot has learned. In practice, however, $\theta \in \mathbb{R}^d$ is a continuous variable, and so the robot's belief over $\theta$ is a \textit{continuous} probability distribution. We therefore introduce $Z \subset \mathbb{R}^m$, a compact, \textit{low-dimensional representation} of the robot's belief space. Here different values $z \in Z$ are associated with different beliefs $b_{\mathcal{R}}$, and the robot's current belief $b_{\mathcal{R}}^i$ is parameterized by $z^*$. To give an example, consider trying to model $b_{\mathcal{R}}^i$ as a normal distribution, where $z^*$ is a vector that contains the mean and standard deviation of this model. In this example $z^*$ is a low-dimensional representation because it takes the continuous probability distribution $b_{\mathcal{R}}^i$ and compactly approximates it with $2d$ parameters.

In the same way that humans cannot directly observe the robot's learning, users cannot directly observe $z^{*}$. But humans can infer a probability distribution over what $z^{*}$ is based on the questions the robot has asked. We refer to this distribution as the \textit{human's belief}:
\begin{equation} \label{eq:P3}
    b_{\mathcal{H}}^{i+1}(z) = P(z \mid Q_1, \ldots, Q_i, q_1, \ldots, q_{i-1})
\end{equation}
Intuitively, $b_{\mathcal{H}}(z)$ expresses how likely it is \textit{from the human's perspective} that the robot has learned $z$ (i.e., $z^{*} = z$) given the robot's questions and the human's previous answers.

\p{Representing Robot Learning} Our proposed approach is not tied to any specific choice of $Z$. But, for the sake of clarity, we here present one representation that we will leverage in our experiments. 

Instead of thinking about the robot's learning directly in terms of the robot's belief $b_{\mathcal{R}}$, we focus on the \textit{behavior} that this belief induces. For any estimate $\theta \sim b_{\mathcal{R}}$ of the human's preferences, the robot can solve for the corresponding trajectory that maximizes $R$:
\begin{equation} \label{eq:P4}
    \xi_{\theta} = \text{arg}\max_{\xi \in \Xi} R(\xi, \theta)
\end{equation}
Here $\xi_{\theta}$ is an optimal trajectory given preferences $\theta$. Imagine that the robot has currently learned $b_{\mathcal{R}}^i$. If we were to deploy this robot now, and the robot solves for the optimal trajectory given what it has learned, the expected features of this trajectory are:
\begin{equation} \label{eq:P5}
    \mu_i = \mathbb{E}_{\theta \sim b_{\mathcal{R}}^i}\big[f(\xi_{\theta})\big]
\end{equation}
Similarly, the variance of these features across multiple runs is:
\begin{equation} \label{eq:P6}
    \sigma_i^2 = \mathrm{Var}_{\theta \sim b_{\mathcal{R}}^i}\big[f(\xi_{\theta})\big]
\end{equation}
Viewed together, Equations~(\ref{eq:P5}) and (\ref{eq:P6}) determine i) what the robot would do if deployed and ii) what features of the task the robot is and is not confident about. We combine both the mean $\mu_i$ and standard deviation $\sigma_i$ to form the vector $z^*$:
\begin{equation} \label{eq:P7}
    z^* = (\mu_i, \sigma_i) \in \mathbb{R}^{2d}
\end{equation}
Under this representation, $z^* \in Z$ is a compact embedding of the \textit{features induced} by the robot's current belief $b_{\mathcal{R}}^i$. Our choice of $Z$ contains both the behavior the robot has learned to perform as well as the parts of this behavior the robot is confused about.

To better understand $z^*$, we return to our motivating example in \fig{front}. Here the task features are \textit{stacking distance} and \textit{height}. If the robot is completely confident it should minimize stacking distance, but is unsure about the right height, the components of $z^*$ could be:
\begin{equation} \label{eq:P8}
    \mu_i = (0 \text{ stacking distance}, 0.5 \text { height}), \quad
    \sigma_i = (0 \text{ stacking distance}, 0.3 \text { height})
\end{equation}
Communicating $z^*$ back to the human has two parts: what the robot has learned (i.e., $\mu_i$) and how confident the robot is in that learning (i.e., $\sigma_i$). An intelligent robot should choose questions to make $z^*$ transparent and bring the human into the learning loop.

%% file: model.tex
\section{Modeling how Humans Interpret Questions} \label{sec:model}

Our work is based on the hypothesis that humans gather information about the robot by observing the questions the robot asks. In this section we formalize our hypothesis by introducing a human model that robots can leverage to solve \eq{P3}. Intuitively, this model defines how humans \textit{interpret} robot questions; more specifically, this model relates the questions the robot asks to the information those questions reveal. A robot that optimizes for this model chooses \textit{revealing} questions that proactively communicate what it has learned.

\p{Bounded Memory} Within Equations~(\ref{eq:P2}) and (\ref{eq:P3}) the human is learning in response to the robot's learning. This is an instance of \textit{co-adaptation}. Prior research from economics \cite{kahneman2013prospect}, game theory \cite{powers2005learning}, and human-robot collaboration \cite{nikolaidis2017human} suggests that co-adaptive humans base their decisions on the recent actions of their partner. Applying this bounded memory model to our setting, we assume that the human focuses on the robot's \textit{last $k$ questions}. We also assume that --- from the human's perspective --- the robot's learned behavior \textit{$z$ is constant} during these $k$ questions. Applying both assumptions, \eq{P3} now reduces to\footnote{Notice that \eq{M1} no longer depends on answers $q$. This is a result of the human's assumption that $z$ is locally constant, meaning that their last few answers did not affect the robot's learning.}:
\begin{equation} \label{eq:M1}
    b_\mathcal{H}^{i+1}(z) = P(z \mid Q_{i-k+1}, \ldots, Q_i)
\end{equation}
Employing Bayes' Theorem and conditional independence:
\begin{equation} \label{eq:M2}
    b_\mathcal{H}^{i+1}(z) \propto P(z) \prod_{\tau = i-k+1}^i P(Q_{\tau} \mid z)
\end{equation}
\eq{M2} captures what the human thinks the robot has learned based on the robot's $k$ most recent questions. Here $P(z)$ is the human's \textit{prior} over what the robot knows, and $P(Q \mid z)$ is the \textit{likelihood} that the robot asks question $Q$ given that it has learned $z$. The last step in our human model is formulating this likelihood function $P(Q \mid z)$.

\p{Question Likelihood} To motivate our choice of likelihood function we return to our running example where the robot is learning about \textit{height} and \textit{stacking distance}. Imagine that the robot shows you a question with two trajectories: one trajectory moves close to the table and stacks the dishes, while the other trajectory moves high above the table and places the dishes in different locations. Because both \textit{height} and \textit{stacking distance} vary in this question, you may infer that the robot is uncertain about both features of the task. By contrast, if the robot shows you a question where both of the trajectories stack the dishes, you will likely assume that the robot knows \textit{stacking distance}. Driven by this intuition, we propose to model $P(Q \mid z)$ by comparing the actual behavior shown in question $Q$ to the learned behavior captured by $z$:
\begin{equation} \label{eq:M3}
    P(Q \mid z) \propto \exp \{- \| (\mu_Q, \sigma_Q) - z \|^2 \}
\end{equation}
Here $d$ is the number of features, $(\mu_Q, \sigma_Q) \in \mathbb{R}^{2d}$ is a vector formed by concatenating $\mu_Q$ and $\sigma_Q$, and $z \in \mathbb{R}^{2d}$ is our parameterization from \eq{P7}. Recall that each question $Q$ is composed of a set of trajectories $Q = \{\xi_1, \ldots, \xi_M\}$. In \eq{M3}, $\mu_{Q}$ is the mean features across the trajectories shown in question $Q$, and $\sigma_{Q}$ is the standard deviation of these features:
\begin{equation} \label{eq:M4}
    \mu_{Q} = \mathbb{E}_{\xi \in Q} \big[ f(\xi) \big], \quad 
    \sigma_{Q}^2 = \mathrm{Var}_{\xi \in Q} \big[ f(\xi) \big]
\end{equation}
In practice, \eq{M3} states that \textit{humans expect the robot to ask questions that match the behavior the robot has learned}. Consider our motivating example where the robot's learned behavior is summarized in \eq{P8}. Under our model, the human thinks it \textit{likely} that this robot will ask questions where each trajectory minimizes stacking distance and demonstrates a different height, but \textit{unlikely} that the robot will ask questions that vary the stacking distance.

\p{Revealing Questions} What if the robot wants to ask questions that purposely convey its learned behavior to the human? Put another way, what if the robot wants to be as \textit{interpretable} as possible? Recall that the robot's current belief $b_{\mathcal{R}}^i$ is compactly represented by $z^*$. We leverage our model from Equations~(\ref{eq:M2})--(\ref{eq:M4}) to proactively reveal $z^*$ to the human:
\begin{equation} \label{eq:M5}
    Q^*_{\text{reveal}} = \text{arg}\max_{Q \in \mathcal{Q}} ~ b_\mathcal{H}^{i+1}(z^*)
\end{equation}
Here $\mathcal{Q}$ is the set of feasible questions, and $Q^*_{reveal} \in \mathcal{Q}$ is the question that greedily maximizes the human's belief in $z^*$. We emphasize that --- because the robot in \eq{M5} is optimizing for transparency --- the human's answer to $Q_{reveal}$ may not be particularly informative for the robot. Instead, this robot is only considering the human's perspective. From the human's point of view $Q_{reveal}$ is interpretable because it demonstrates a set of trajectories which accurately reflect what the robot is confident about and what the robot is still trying to learn.

\p{Summary} With Equations~(\ref{eq:M2})--(\ref{eq:M4}) we formalize our key insight: we introduce a cognitive model that relates the questions that the robot asks to the information those questions reveal. As shown in \eq{M5}, this model enables robots to proactively optimize for revealing questions.

%% file: amt.tex
\section{Do Humans Learn from Questions?} \label{sec:amt}

\begin{figure*}[t]
	\begin{center}
		\includegraphics[width=1\columnwidth]{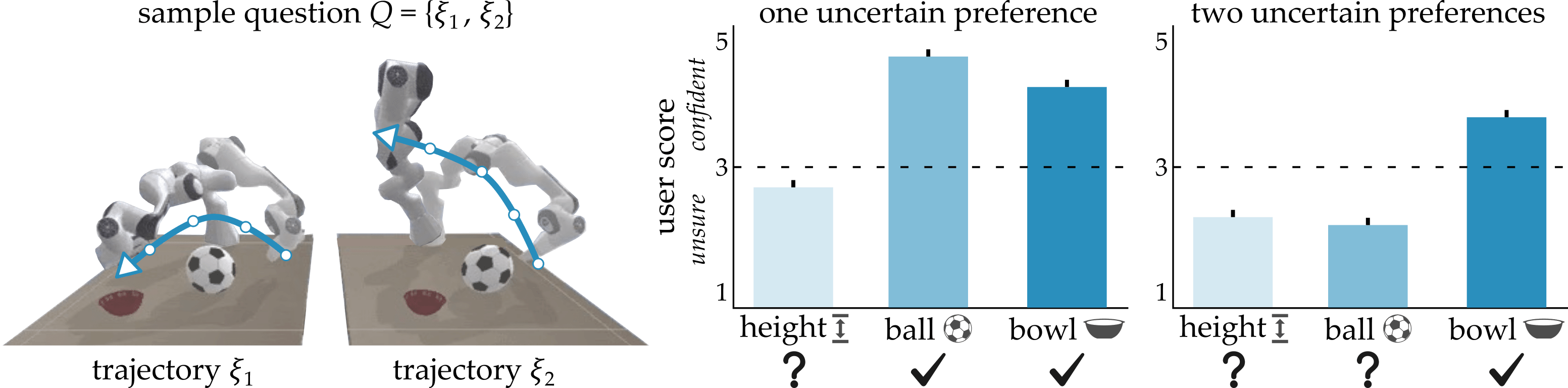}

		\caption{Questions reveal information to humans. (Left) An example question from our Amazon Mechanical Turk user study. (Right) Participants responded to sequences of these robot questions by indicating what they thought the robot had learned. High scores indicate the human believed the robot was confident about a preference, while low scores indicate the human perceived the robot as unsure about that preference. In reality, we displayed i) robots that were only unsure about \textit{height} and ii) robots that were unsure about both \textit{height} and \textit{ball}. Across both conditions users correctly inferred what the robot had learned based only on the questions the robot asked. Differences in user score were statistically significant $(p<.001)$.}
		
		\label{fig:amt1}
	\end{center}

\end{figure*}

We have proposed a model to formalize the relationship between robot questions and the information they convey. Here we conduct an online user study to test this model and explore whether humans gather information from robot questions. Within this study a robot arm moves across a table to learn three human preferences (see \fig{amt1}): the \textit{height} of the robot's end-effector, the distance between the robot and the \textit{ball}, and whether or not the robot ends above the \textit{bowl}. The robot asks the human a sequence of questions to learn their preferences. \textit{We measure what these questions reveal to the human}. After viewing the questions, participants specify which preferences they think the robot understands or is unsure about. The robot optimizes for questions that are as revealing as possible: if humans follow our cognitive model, we anticipate that their understanding of what the robot knows will match what the robot has \textit{actually} learned.

\p{Independent Variables} We varied the robot's uncertainty across two scenarios. In the first scenario the robot was unsure about \textit{height} but confident about \textit{ball} and \textit{bowl}. In the second scenario the robot was unsure about both \textit{height} and \textit{ball} while remaining confident about \textit{bowl}. We never informed our study participants which preferences the robot did and did not know.

\p{Participants and Procedure} We conducted a within-subjects study on Amazon Mechanical Turk and recruited $50$ participants. All participants had at least a $99\%$ approval rating. We first described the high-level task, and explained the \textit{height}, \textit{ball}, and \textit{bowl} preferences. We then asked comprehension questions to ensure that all participants had carefully read and understood our instructions. The remainder of the survey was divided into six sections: each section included videos of three robot questions (similar to the sample question from \fig{amt1}). In half of the sections the robot was uncertain about one feature, and in half of the sections the robot was uncertain about two features. We randomized the order of the sections.

The robot leveraged our cognitive model and \eq{M5} to choose which questions to ask. This \textit{revealing} robot optimized for questions that communicate robot learning to the human --- if users were unable to interpret these questions, we doubted users would gather information from other, more ambiguous questions. 

\p{Dependent Measures} After viewing each set of robot questions, participants indicated what they thought the robot had learned on a $5$-point scale. Here a score of $1$ denotes that the human believed the robot was completely unsure about a given preference, while a score of $5$ denotes that the human thought the robot fully understood that preference. We aggregated all $50$ user responses to determine the mean score and standard error of the mean (SEM) for each preference.

\p{Hypothesis} \textit{When robots purposely ask revealing questions, humans will correctly infer what parts of the task the robot is confident or unsure about.}

\p{Results} Our results are visualized in \fig{amt1}. When interpreting these results, remember that a user score \textit{below} $3$ indicates participants thought the robot was unsure about a preference, while a score \textit{above} $3$ indicates humans thought the robot was confident. Overall, we found that participants correctly inferred what the robot did and did not know based just on the questions the robot asked.

\textit{-- One uncertain preference.} When the robot was only uncertain about \textit{height}, post hoc comparisons revealed that users perceived the robot as more unsure of \textit{height} than either \textit{ball} or \textit{bowl} ($p < .001$).

\textit{-- Two uncertain preferences.} When the robot was unsure about both \textit{height} and \textit{ball}, post hoc comparisons demonstrated that participants thought the robot was more uncertain about \textit{height} and \textit{ball} as compared to \textit{bowl} ($p < .001$).

\begin{figure}[t]
	\begin{center}
		\includegraphics[width=0.5\columnwidth]{Figures/amt_2.pdf}

		\caption{Follow-up comparison between two types of revealing questions. In \textbf{all-at-once} the robot asks questions that reveal what the robot knows about every preference, while in \textbf{one-at-a-time} the robot asks questions that reveal a single preference per question. The robot is actually uncertain about \textit{height} and \textit{ball} preferences. We found that differences between the participants' perception of \textbf{all-at-once} and \textbf{one-at-a-time} were not statistically significant.}
		
		\label{fig:amt2}
	\end{center}

\end{figure}

\smallskip \noindent \textbf{Do Human's Focus on Specific Parts of Questions?} When conducting this study we wondered whether humans viewed robot questions holistically, or if they focused their attention on specific aspects of the robot's questions. Consider the example in \fig{amt1}. Here the robot's question shows two clearly different \textit{heights}, but also two slightly different values for \textit{bowl}: $\xi_1$ ends directly above the bowl, while $\xi_2$ is a bit to the left of the bowl. When participants saw this question, did they only learn about \textit{height}, or did they also think about \textit{bowl}?

Similar to prior work \cite{bajcsy2018learning}, we therefore performed a follow-up user study with \textbf{all-at-once} and \textbf{one-at-a-time} questions. We conducted this survey within the scenario where the robot was unsure about two preferences: \textit{height} and \textit{ball}. The \textbf{all-at-once} robot asked questions that conveyed what it had learned about every preference simultaneously (e.g., a question that varied both \textit{height} and \textit{ball}). By contrast, the \textbf{one-at-a-time} robot only conveyed its understanding of a single preference per question (e.g., the first question only varied \textit{height}, and the second questions only varied \textit{ball}). We recruited $50$ users on Amazon Mechanical Turk for this follow-up study.

Our results are shown in \fig{amt2}. Interestingly, participants interpreted \textbf{all-at-once} questions just as well as they interpreted \textbf{one-at-a-time} questions: the differences between these two conditions were not statistically significant ($F(1,49)=0.20$, $p=.66$). This suggests that participants in our survey did not hone in on specific parts of robot questions, and robots could leverage a single question to convey multiple aspects of their learning. We recognize that a limitation of this study was that there were only three total preferences to ask about: we anticipate that as the number of preferences increases, the \textbf{all-at-once} approach may break down.

\p{Summary} The key assumption underlining our work is that robot questions reveal information to humans. We supported this hypothesis in an online user study: $50$ participants correctly identified what the robot did and did not know based only on the questions the robot asked. Within this study the robot purposely optimized for \textit{revealing} questions using our cognitive human model from \eq{M5}. The effectiveness of these questions suggests that our model produces questions that the human can correctly interpret.

%% file: questiontypes.tex
\section{Combining Informative and Revealing Questions} \label{sec:info}

In the previous section we demonstrated that robot questions can convey information to human onlookers. We generated these questions using \eq{M5}, which optimizes for \textit{revealing} questions. On the one hand, revealing questions communicate what the robot has learned \textit{to the human}. On the other hand, these questions may not be particularly informative \textit{for the robot}.

To see why, we return to our motivating example in \fig{front}. Here the robot is unsure about how high to carry the dishes, and so when it optimizes for a revealing question, it shows trajectories that carry the dishes at fluctuating heights. Perhaps one trajectory initially moves the dishes close to the table and then goes high, while the second trajectory initially moves high and then goes low. This question conveys to the human that the robot is unsure about height (since each $\xi \in Q$ moves at varying height), but the human's answer to this question provides \textit{no new information} to the robot (because each $\xi \in Q$ has the same average height). From the robot's perspective, we want to ask a question that clearly elicits the human's preference: e.g., one trajectory that carries the dishes high above the table, and a second trajectory that always keeps the dishes near the table.

In this section we formulate revealing questions as one extreme of a spectrum. At the other end of this spectrum are \textit{informative} questions that maximize the information gained by the robot. Informative questions speed up robot learning but may mislead the human; revealing questions clarify what the robot has learned but may not elicit new information. We describe how robots trade-off along this spectrum, intelligently balancing between giving and getting information.

\p{Informative Questions} Optimizing for informative questions is well explored in prior work \cite{sadigh2017active, biyik2020learning, ibarz2018reward}. Following \cite{biyik2019asking, biyik2020learning}, informative questions greedily maximize the robot's expected information gain about $\theta^*$, the human's true preferences:
\begin{align} \label{eq:T1}
\begin{split}
    Q^*_{\text{info}} & = \text{arg}\max_{Q \in \mathcal{Q}} ~ I(\theta^* ; q \mid Q, b_{\mathcal{R}}^i) \\
    & = \text{arg}\max_{Q \in \mathcal{Q}} ~ H(\theta^* \mid Q, b_{\mathcal{R}}^i) - \mathbb{E}_{q \sim Q} \big[H(\theta^* \mid q, Q, b_{\mathcal{R}}^i) \big]
\end{split}
\end{align}
Here $I$ denotes mutual information, $H$ is the Shannon entropy, and $I(\theta^* ; q \mid Q, b_{\mathcal{R}}^i)$ quantifies the amount of information the robot will learn about $\theta^*$ if it observes human answer $q$ given question $Q$ and current estimate $b_{\mathcal{R}}^i$ \cite{cover2012elements}. The robot does not know the human's answer $q$ when choosing which question to ask. Accordingly, in \eq{T1} the robot chooses its question $Q_{\text{info}}^* \in \mathcal{Q}$ such that --- regardless of what answer the human provides --- observing that answer will optimize the expected amount of information the robot gains.

\p{Trading-Off} A robot that optimizes \eq{T1} focuses on purely \textit{gaining information} from the human. Conversely, a robot that optimizes \eq{M5} is only focused on \textit{revealing information} to the human. In order to facilitate a bi-directional exchange of information, we here introduce a spectrum between these two extremes. Robots can now identify questions that are both informative and revealing by optimizing:
\begin{equation} \label{eq:T2}
    Q^*_{\text{spectrum}} = \text{arg}\max_{Q \in \mathcal{Q}} ~ I(\theta^* ; q \mid Q, b_{\mathcal{R}}^i) + \lambda \cdot b_\mathcal{H}^{i+1}(z^*)
\end{equation}
In the above, $\lambda \geq 0$ is a hyperparameter that determines the relative importance of gaining or revealing information. When $\lambda = 0$ the robot only considers information gain, and if $\lambda \rightarrow \infty$ the robot thinks only about revealing information.

\smallskip

\noindent\textbf{When Should Robots Gather or Reveal Information?} Robots that leverage our approach trade-off along a spectrum between getting and giving information. We want to theoretically understand when \eq{T2} leads to more informative questions or more revealing questions. Below we prove that robots which leverage \eq{T2} follow an underlying pattern: they ask their most informative questions the first time they interact with the human, and converge towards purely revealing questions over repeated interactions.

\smallskip

\noindent \textbf{Proposition 1.} 
\textit{Assuming that the user has fixed preferences $\theta^*$, and the user chooses answers that are consistent with their preferences, robots that optimize \eq{T2} for informative and revealing questions converge towards asking revealing questions over time}.

\smallskip

\noindent \textit{Proof.} We assume that each time the robot asks a question the human chooses a trajectory $q \in Q$ that maximizes their reward, so that $q \in \text{arg}\max_{q' \in \mathcal{Q}} ~R(q', \theta^*)$. Given this assumption, the first term from \eq{T2} is submodular and monotone \cite{nemhauser1978analysis}. It directly follows that $I(\theta^* ; q \mid Q, b_{\mathcal{R}}^i) \rightarrow 0$ as the interaction number increases (i.e., $i \rightarrow \infty$). Intuitively, each successive question and answer provides less new information about $\theta^*$ since the robot has already formed an increasingly accurate estimate from the previous human answers. At the limit $I(\theta^* ; q \mid Q, b_{\mathcal{R}}^i) \rightarrow 0$, and the robot only optimizes for $b_\mathcal{H}^{i+1}(z^*)$, i.e., revealing questions. \qed

\p{Implementation and Hyperparameter Tuning} When a robot leveraging \eq{T2} first interacts with the human it does not know what the human wants, and selects questions that proactively learn $\theta^*$. Over time this robot gets a good grasp of the human's preferences, so that there is no longer any need to ask informative questions. At this point the robot continuously transitions towards revealing questions that communicate $z^*$ back to the human.

There are two hyperparameters for designers to tune when implementing \eq{T2}. The first is $\lambda$, which arbitrates the relative importance of informative and revealing questions. The second is $k$ from \eq{M2}, which models how many recent questions the human remembers. We will study the effects of both hyperparameters in the following sections. Based on these experimental results, we recommend that designers start with $\lambda=1$ and $k \geq 3$. We find that that optimizing for informative and revealing questions is largely \textit{invariant} to the choice of $k$, while $\lambda$ enables the designer to adjust how informative or revealing the robot is.

%% file: simulations.tex
\section{Simulations}
\label{sec:sim}

We have formulated informative and revealing questions along a continuous spectrum, and proposed that robots should trade-off between these extremes. However, it is not yet clear whether robots that optimize for \textit{both} informative and revealing questions will do a good job of either one. Put another way, when the robot optimizes for \eq{T2}, are its questions rich enough to gain information while remaining transparent to the human onlooker? Here we test our proposed approach in two environments with simulated users, and compare informative and revealing questions to the performance of state-of-the-art active learning baselines. We then conduct two separate simulations where we vary the hyperparameters in our approach: we experimentally find that humans do not need to remember every robot question in order to correctly interpret the robot's learning.

\p{Baselines} We compare the performance of informative and revealing questions (\textbf{Ours}) to three different baselines. The first is a na\"ive robot that chooses questions uniformly at \textbf{Random}. The other two baselines are the extremes of our spectrum: one robot optimizes solely for \textbf{Informative} questions using \eq{T1}, and another robot optimizes for \textbf{Revealing} questions using \eq{M5}. We note that the revealing robot matches what we used in our online user study from Section~\ref{sec:amt}, and the informative robot is taken from recent work on active preference learning \cite{biyik2019asking, biyik2020learning}. Hence, \textbf{Informative} is a \textit{state-of-the-art} approach for active preference-based reward learning.

\p{Simulated Users} For each robot question we simulate the human's response. There are two components of this response: i) what the human infers from the robot's question and ii) how the human answers this question. To simulate what the human infers from the robot's question we leverage the cognitive human model we developed in Section~\ref{sec:model} and experimentally tested in Section~\ref{sec:amt}. To simulate how the human answers the question we apply the same approach as related work on active learning \cite{holladay2016active, basu2018learning, sadigh2017active, ibarz2018reward, biyik2020learning}. Here the human is modeled as an approximately optimal decision maker. In each question the robot shows two different trajectories, $Q = \{\xi_1, \xi_2\}$, and the human has three possible answers: trajectory $\xi_1$, $\xi_2$, or ``I don't know'' \cite{holladay2016active}. The human noisily chooses the trajectory $\xi$ that best agrees with their preferences, and in cases where the two trajectories produce about equal rewards, the human is more likely to select ``I don't know'' (Idk):
\begin{equation*}
    P(q=\xi_i \mid Q, \theta) = \frac{\exp(R(\xi_i, \theta))}{\exp(R(\xi_i, \theta)) + \exp(1 + R(\xi_{-i}, \theta))}
\end{equation*}
\begin{equation*}
    P(q=\text{Idk} \mid Q, \theta) =
    P(q=\xi_1 \mid Q, \theta) \cdot P(q=\xi_2 \mid Q, \theta) \cdot (\exp(2) - 1)
\end{equation*}
In the above we use $-i$ to denote the other trajectory, so that if $i=1$ then $-i = 2$ (and vice versa). Our choice of human model is consistent with prior work on inverse reinforcement learning \cite{ziebart2008maximum, jeon2020reward, osa2018algorithmic} and cognitive psychology \cite{luce2012individual, krishnan1977incorporating, kahneman2013prospect}.

\p{Environments} We implement the simulated users alongside a robot arm and autonomous car \cite{coumans2016pybullet}. The robot arm environment matches our previous user study (see \fig{amt1}), where the human's unknown preferences are the robot's height, the distance to the ball, and if the robot ends above the bowl. In the driving environment the human's preferences include the car's speed, the car's distance from obstacles, and the car's distance from the center of the lane. Across both environments we simulate $100$ users with uniformly randomly chosen preferences $\theta^*$.

\begin{figure*}[t]
	\begin{center}
		\includegraphics[width=1\columnwidth]{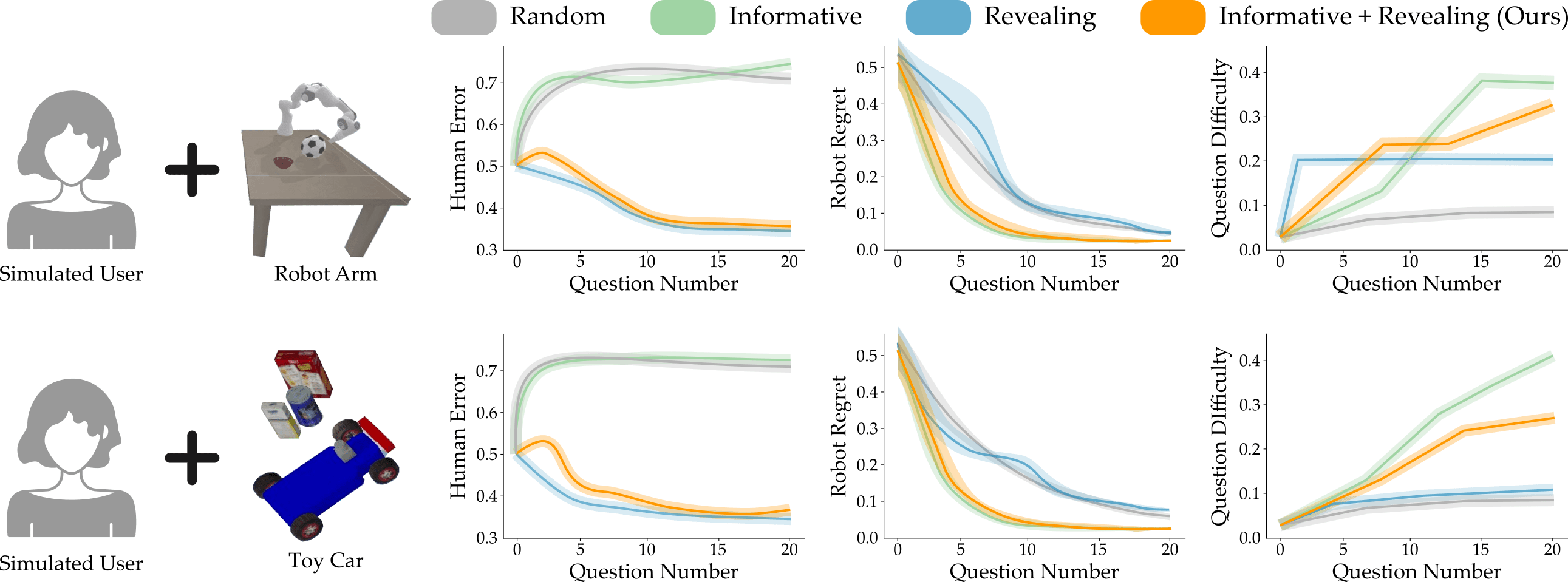}

		\caption{Simulated humans answering $20$ robot questions. \textit{Human Error} refers to the difference between what the robot actually learned and what the human thinks the robot learned (i.e., the human's perspective). By contrast, \textit{Robot Regret} captures the error between the robot's learned behavior and the human's desired behavior (i.e., the robot's perspective). When we consider both the robot and human perspectives we generate questions that convey learning while gathering information (\textbf{Ours}). \textit{Question Difficulty} denotes the fraction of questions in which the simulated human did not know how to answer and selected ``I don't know.'' Humans find questions generated by our approach less challenging to answer than informative questions. Note that lower is better in all plots. Shaded regions show standard deviation across $100$ simulated humans.}
		
		\label{fig:sims}
	\end{center}

\end{figure*}

\subsection{Comparison to State-of-the-Art Baselines}

Our first set of simulations compares informative and revealing questions to na\"ive and state-of-the-art baselines. The results from these simulations are visualized in \fig{sims}.

\smallskip
\noindent\textbf{Are Questions Revealing?} We first consider the human's perspective, and determine which types of questions communicate the robot's learning back to the human. We measure \textit{human error}, which captures the difference between what the robot really knows (i.e., $z^*$) and what the human thinks the robot knows (i.e., $b_{\mathcal{H}}$). This is a proxy measure of \textit{interpretability}: ideally, the human will understand exactly what their robot has learned so that they can correctly anticipate what their robot will do when it is deployed. Looking at \fig{sims}, we see that both \textbf{Ours} and \textbf{Revealing} are transparent to the human. By contrast, simulated users who interact with either \textbf{Random} or \textbf{Informative} are unsure about what the robot partner learned. Indeed, \textbf{Informative} questions tell the human just as much about what the robot is learning as completely \textbf{Random} questions do.

\smallskip
\noindent\textbf{Are Questions Informative?} We next consider the robot's perspective, and determine which types of questions provide the most information about the human's latent preferences. We assess how quickly the robot learns by measuring \textit{regret} \cite{osa2018algorithmic}. Recall from \eq{P4} that $\xi_{\theta}$ is the optimal trajectory given that the human's preferences are $\theta$. Regret is the difference in reward between the human's preferred behavior and the robot's learned behavior: $\mathbb{E}_{\theta \sim b_{\mathcal{R}}}\big[R(\xi_{\theta^*}, \theta^*) - R(\xi_{\theta}, \theta^*)\big]$. Regret will decrease over time because the robot is collecting more human answers and honing in on the human's preferred behavior $\theta^*$. But we can \textit{accelerate} this process by proactively asking insightful questions --- referring to \fig{sims}, both \textbf{Our} robot and the \textbf{Informative} robot quickly learned the human's desired behavior. By contrast, a robot that asks \textbf{Revealing} questions learned at about the same rate as a robot that was asking questions completely at \textbf{Random}.

\smallskip
\noindent\textbf{Are Questions Confusing?} Beyond both gaining and revealing information we want to ensure that the robot's questions are \textit{easy} for the human to answer. Prior work suggests that easy, user-friendly questions improve the human's experience and reduce the number of wrong answers, which in turn speeds up the robot's learning \cite{biyik2019asking, biyik2020learning, cakmak2012designing, racca2019teacher}. We therefore assess \textit{question difficulty}, which measures the fraction of time that the simulated humans answered "I don't know." When users pick "I don't know" the options are too hard to distinguish and the question is challenging for the human to clearly answer. In general, we find that \textbf{Random} and \textbf{Revealing} questions are quite easy for the human to answer, while \textbf{Informative} and \textbf{Ours} get harder over time. \textbf{Ours} is slightly easier than the \textbf{Informative} baseline.

\subsection{Effect of Hyperparameters on Our Approach}

\begin{figure*}[t]
	\begin{center}
		\includegraphics[width=1\columnwidth]{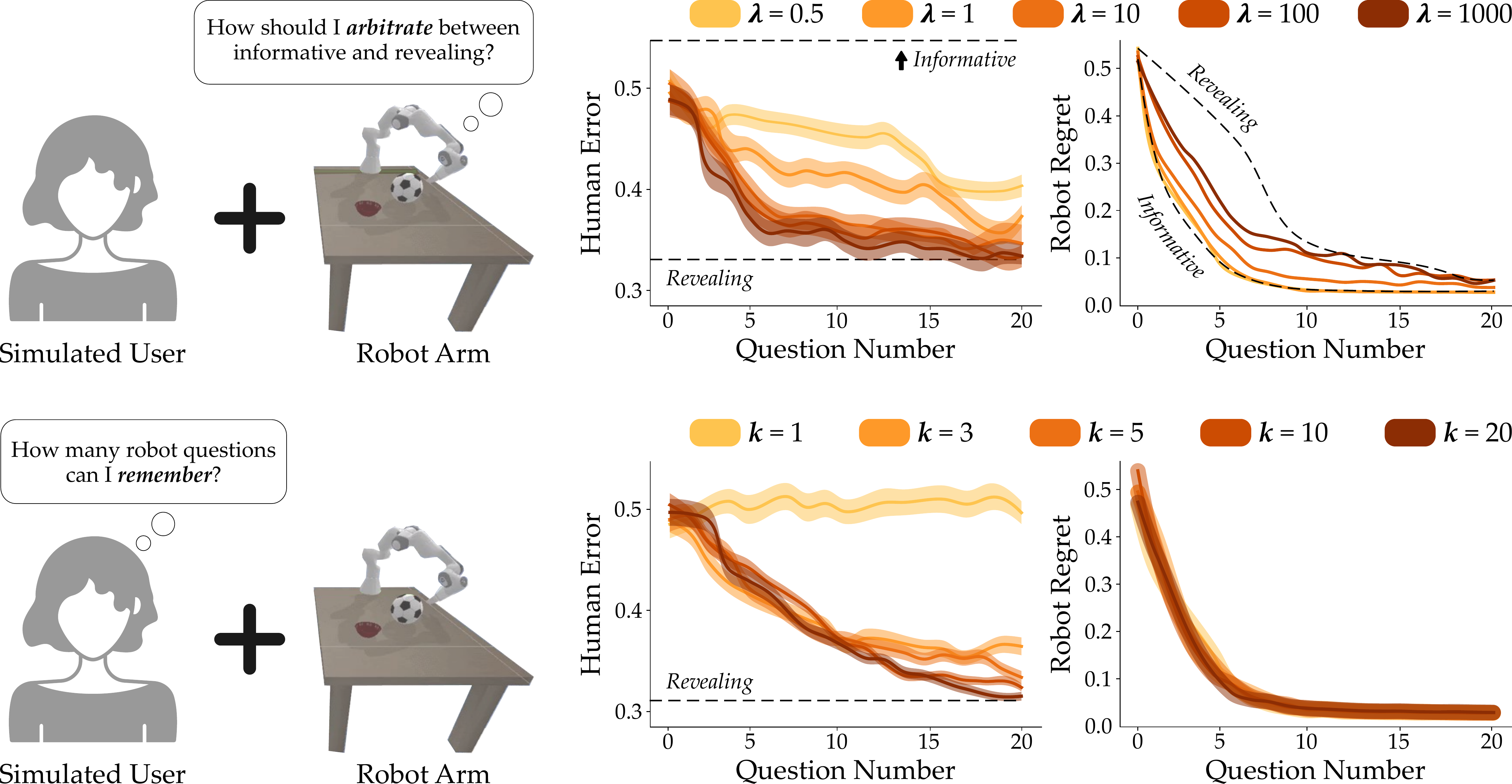}

		\caption{Simulated humans interacting with our proposed approach. We tune two key hyperparameters and see how they affect the human's understanding of the robot (\textit{Human Error}) and the robot's estimate of the human's preferences (\textit{Robot Regret}). The dashed lines show the lower and upper bounds for revealing or informative baselines. (Top) We leave $k=3$ while varying the relative importance of gaining and revealing information. At high values of $\lambda$ the robot favors questions that reveal information, and at low values of $\lambda$ the robot focuses on gaining information. Even at the extremes ($\lambda = 0.5$, $\lambda = 1000$) our combined approach performs better than just asking informative or revealing questions. (Bottom) We leave $\lambda = 1$ and vary how many recent questions the simulated human reasons over. The more questions the human remembers the more information the robot can reveal; however, our approach still reaches the lower bound even if the human can only remember the last $3$ questions. Standard deviation omitted from the top right for clarity.}
		
		\label{fig:sims-params}
	\end{center}

\end{figure*}

In the previous simulations we choose $\lambda = 1$ and $k=3$ for the hyperparameters of \textbf{Ours} and left these hyperparameters fixed across both environments. Now in our second set of simulations we will explore how tuning these two hyperparameters affects our approach to generating informative and revealing questions. The results from these simulations are visualized in \fig{sims-params}.

\smallskip
\noindent\textbf{How Should Designers Arbitrate?} In \eq{T2} we introduce $\lambda$, which arbitrates the relative importance of gaining or revealing information. Higher values of $\lambda$ put more weight on revealing information, while lower put more weight on gaining information. In \fig{sims-params} we return to the simulated robot arm environment, and test \textbf{Our} approach with $\lambda = \{0.5, 1, 10, 100, 1000\}$. We compare these results to the lower and upper bounds for \textbf{Informative} and \textbf{Revealing} robots. We find that $\lambda$ effectively enables designers to tune between these extremes: at low values of $\lambda$ we approach the lower bound for robot regret, and at high values of $\lambda$ we approach the lower bound for human error. Importantly, even at $\lambda = 0.5$ the robot is still asking questions that are more revealing than an \textbf{Informative} robot. This suggests that --- even when designers are focused on quickly learning human preferences --- \textbf{Ours} can still noticeably improve human understanding.

\smallskip
\noindent\textbf{How Many Questions Does the Human Need to Remember?} In our cognitive human model from \eq{M2} we assume that the human remembers and reasons over the robot's last $k$ questions. In our first set of simulations the human only remembered $k=3$ recent questions: can the robot still convey its learning if the human only thinks about the robot's last question, and how much does this communication improve if the human remembers every question? In \fig{sims-params} we compare \textbf{Our} approach with $k = \{1, 3, 4, 10, 20\}$, where $k=1$ means that the human only remembers the last question, and $k=20$ indicates that the human remembers every question. As expected, the human's bounded memory has no impact on the robot's regret, since $k$ does not change how the human answers the robot's questions. But $k$ does affect how accurately the human interprets the robot's learning. Specifically, we find that a single question is not enough: with $k = 1$ the human's error remains constant after each iteration, while with $k \geq 3$ the human's error converges to the best case performance of a purely \textbf{Informative} robot. We conclude that our approach requires that users keep multiple questions in mind, but they do not need to remember \textit{all} of the questions --- here it was sufficient to keep track of the last three.

\p{Summary} Overall, we find that optimizing for informative and revealing questions captures the best aspects of both perspectives. When using our approach the robot learns at the same rate as the informative baseline while communicating just as well as a revealing robot (\fig{sims}). The questions we generate are still easy for the human to answer: simulated users choose ``I don't know'' less frequently with our method than they do with an existing approach for asking easy questions \cite{biyik2019asking}. Finally, our approach is robust to different choices of the bounded memory model: the human can infer what the robot is learning even if they cannot remember every robot question (\fig{sims-params}).

%% file: userstudy.tex
\section{User Studies}

We conducted two in-person user studies to compare our approach to a state-of-the-art active learning baseline. Users interacted with a robot optimizing purely for informative questions, as well as a robot optimizing for both informative and revealing questions. In the first user study we explored how the robot learned the participants' preferences and the types of questions the robot asked. Next, in our follow-up user study we evaluated how the robot's questions affected the participants' perception of i) what the robot learned and ii) whether it was ready to be deployed. Both studies shared a common experimental setup where participants taught a 7-DoF robot arm (Fetch, Fetch Robotics) by answering the robot's questions across three tasks. Videos of our experimental setup and results are available here: \url{https://youtu.be/tC6y_jHN7Vw}.

\p{Independent Variables} We tested two different algorithms for generating questions: \textbf{Informative} and \textbf{Ours}. In \textbf{Informative}, the robot leveraged a \textit{state-of-the-art} active learning approach that optimizes for information gain \cite{biyik2019asking, biyik2020learning, sadigh2017active}. This robot only seeks to gather information, and does not consider how its own behavior could influence the human. We experimentally compared this state-of-the-art approach to a robot asking informative and revealing questions (\textbf{Ours}). Here the robot applied \eq{T2} to trade-off between both human and robot perspectives.

\subsection{Learning the Human's Preferences} \label{sec:user1}

Our online survey from Section \ref{sec:amt} demonstrated that humans can infer what the robot is learning based on the questions the robot asks. Our first in-person user study goes one step farther: we test whether robots can actively \textit{elicit} information from the human while \textit{also revealing} what they have learned. Participants taught the robot their preferences by answering pairwise comparisons. We measured how efficiently the robot learned the human's preference, and also compared the types of questions the robot asked.

\begin{figure*}[t]
	\begin{center}
		\includegraphics[width=1\columnwidth]{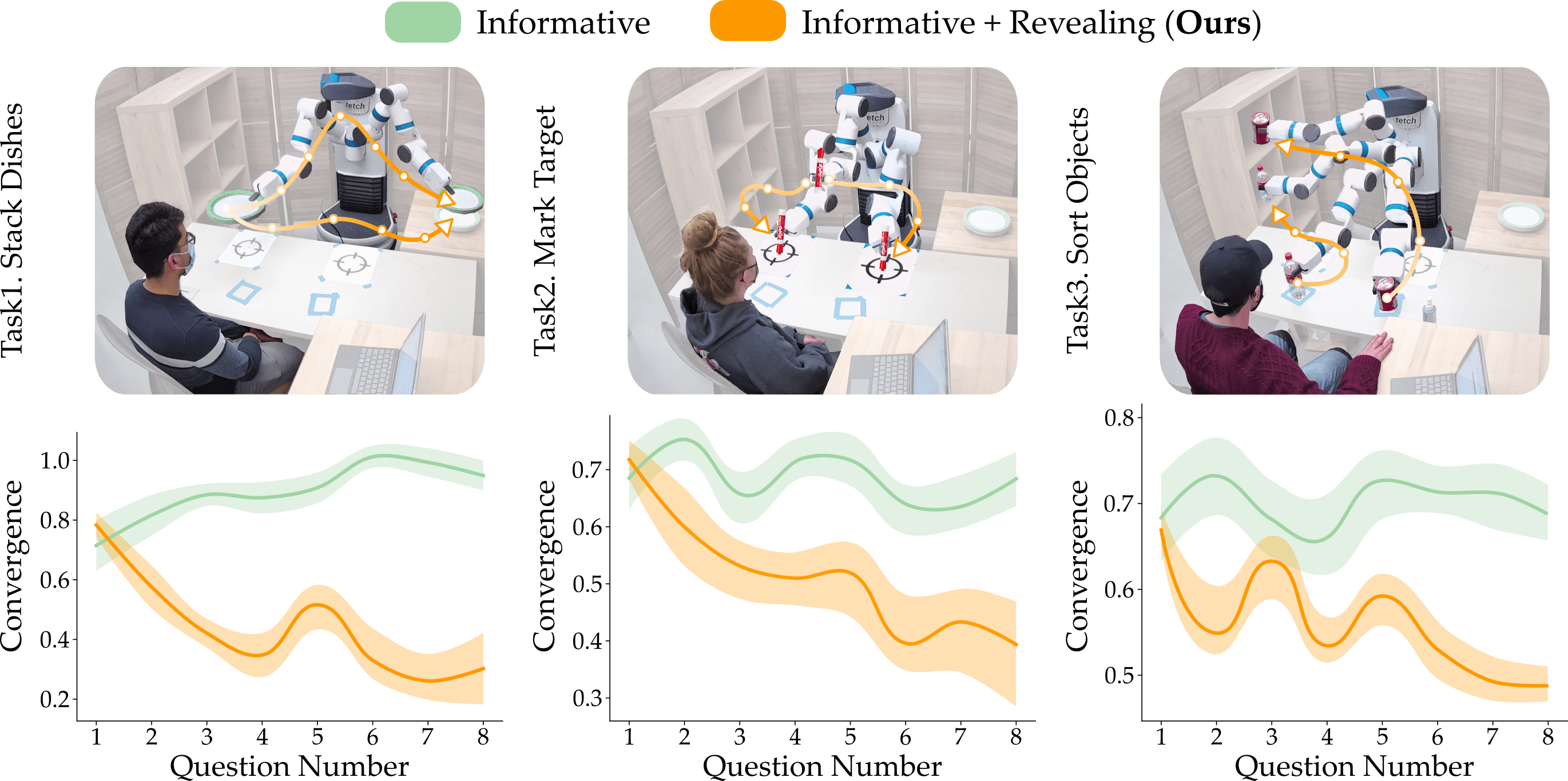}
		\caption{Experimental setup for our user study in Section~\ref{sec:user1}. (Top) The robot asked questions to learn the human's preferences $\theta^*$ across three different tasks. We plot sample question trajectories generated by \textbf{Ours}. (Bottom) When the robot considers how the human interprets its questions, the robot purposely chooses queries that reveal its current understanding $z^*$. Here \textit{Convergence} captures the difference between what the robot learned and the questions the robot asked. With our approach convergence decreases over time, indicating that the robot is asking revealing questions that reflect what it has learned.}
		\label{fig:user_study}
	\end{center}

\end{figure*}

\p{Participants and Procedure} We recruited 10 community members (3 female, 1 non-binary, ages $23 \pm 5.2 $ years) to participate in our study. All participants provided informed written consent consistent with Virginia Tech IRB \#20-755. Participants taught a robot the three tasks shown in \fig{user_study}: stacking dishes (\textit{Stack}), marking a target (\textit{Mark}), and sorting objects (\textit{Sort}). In each task the robot was initially unsure about the human's true preference $\theta^*$. \textit{Stack} is the task from our motivating example (also see \fig{front}) where the robot must learn where to place the dishes and how high to carry them. In \textit{Mark} the robot did not know how and where to mark each target on the table, and in \textit{Sort} the robot was unsure about which object should be placed on which shelf. We leveraged a counterbalanced, within-subjects design: participants completed each task twice, once with \textbf{Informative} and once with \textbf{Ours}. Half of the participants started with the \textbf{Informative} robot. Participants were never told which condition they were interacting with.

During a given interaction the robot asked a total of eight questions, where each question consisted of two possible trajectories ($\xi_1$ and $\xi_2$). Similar to our simulation setup from Section \ref{sec:sim}, participants responded to the robot's question by indicating that they preferred $\xi_1$, $\xi_2$, or by saying ``I don't know.'' At the end of these eight questions the robot played back the optimal trajectory it had learned from the human's answers.

\p{Dependent Measures -- Subjective} After participants answered all the robot's questions with a given condition we administered a 7-point Likert scale survey \cite{schrum2020four}. Our survey questions were arranged into the five multi-item scales described in Table~\ref{table:likert}. We first asked participants how \textit{Easy} it was to answer the robot's questions, whether the robot learned over time (\textit{Improve}), to what extent the robot questions \textit{Revealed} information, and whether it was clear what the robot was unsure about (\textit{Focus}). After the participants answered these initial survey questions we then played back the optimal trajectory that the robot had learned. Participants watched the trajectory, and assessed whether this \textit{Learned Behavior} matched what they originally wanted the robot to do.

\p{Dependent Measures -- Objective} To get a sense of what types of questions the robot chose we measured \textit{Convergence}. Let $\xi^*$ be the trajectory that optimizes the reward function learned from the human's answers, and recall that $Q = \{\xi_1, \xi_2\}$. Convergence captures the feature difference between the robot's current question and the learned trajectory: $\|f(\xi^*) - f(\xi_1)\| + \|f(\xi^*) - f(\xi_2)\|$. In practice, the lower the convergence value gets, the more similar the robot's questions are to the robot's learned behavior (i.e., $\xi_1$ and $\xi_2$ match $\xi^*$). Measuring convergence tells us whether the robot is asking questions to gather new information (higher convergence values) or to reveal what it has learned (lower convergence values).

\p{Hypotheses} We had two hypotheses in this user study:
\begin{displayquote}
    \textbf{H1.} \emph{Robots that trade-off between informative and revealing questions will learn at a similar rate to informative robots while updating the human about the robot's learning.} 
\end{displayquote}
\begin{displayquote}
    \textbf{H2.} \emph{Taking the human's perspective into account will cause the robot to ask questions that converge towards its learned behavior.} 
\end{displayquote}

\begin{figure}[h]
	\begin{center}
		\includegraphics[width=0.8\columnwidth]{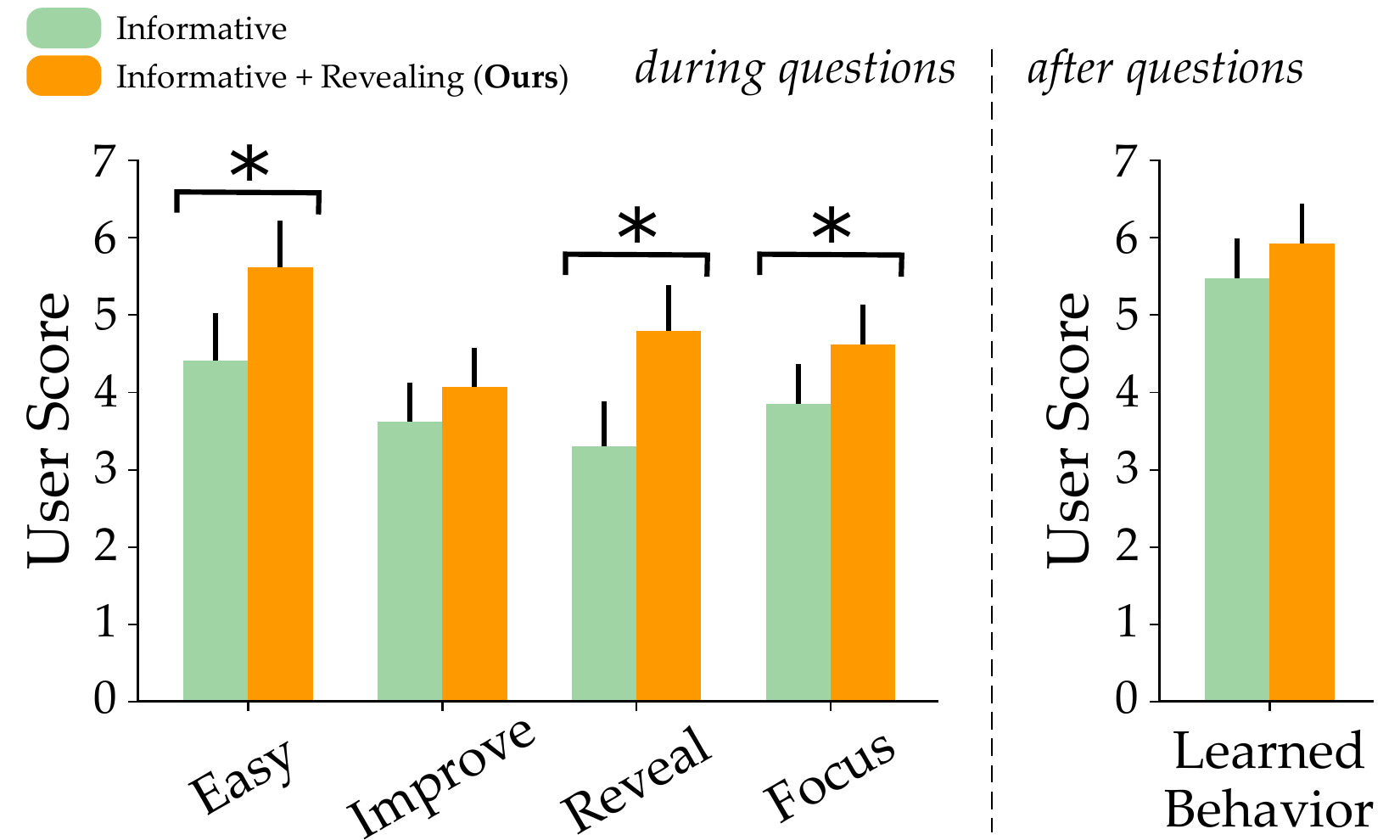}
		\caption{Subjective results from our in-person user study. Higher ratings indicate agreement, and $*$ denotes statistical significance ($p < .01$). (Left) Participants found the questions asked by \textbf{Ours} to be easier to answer, more revealing, and better focused on specific parts of the task. (Right) After asking questions and collecting the user's answers, the robot played back its learned behavior. Participants did not notice a significant difference between what the robot learned with \textbf{Ours} and an \textbf{Informative} active learning approach.}
		\label{fig:subjective}
	\end{center}

\end{figure}

\p{Results -- Subjective} Table~\ref{table:likert} and \fig{subjective} display the results of our Likert scale survey. We first tested the reliability of our five scales, and found that the easy, reveal, focus, and learned behavior scales were reliable (Cronbach's $\alpha > 0.7$). Accordingly, we grouped each of these scales into a combined score, and performed a one-way repeated measures ANOVA on the results. The answers to the \textit{improve} questions were inconsistent, perhaps because different users interpreted these survey questions in different ways. We have included \textit{improve} in our plot of the results, but because of its unreliability we did not conduct additional statistical analysis on this scale.

Overall, the results from our Likert scale survey support \textbf{H1}. Participants perceived the final behavior learned by \textbf{Ours} as comparable to the behavior learned by the \textbf{Informative} baseline. Both approaches successfully learned the human's preferences across all three tasks ($ F(1, 78) = 1.44, ~ p = .233$). However, the questions asked by \textbf{Ours} had a significant effect on participants' attitudes toward teaching the robot. Specifically, users thought the questions asked by \textbf{Ours} were easier to answer ($ F(1, 78) = 11.12, ~ p < .01$), more revealing about the robot's learning ($ F(1, 78) = 27.81, ~ p< .01$), and better focused on uncertain aspects of the task ($ F(1, 78) = 10.55, ~ p < .01$). We conclude that \textbf{Ours} not only learned as efficiently as the state-of-the-art \textbf{Informative} robot, but our approach also revealed this learning back to the human user.

\begin{table*}[t]

	\caption{Questions on our Likert scale survey. We grouped questions into five scales and tested their reliability using Cronbach's $\alpha$. We explored whether the robot asked easy questions, improved over time, revealed information about its learning, focused on specific parts of the task, and learned the correct behavior. Computed $p$-values indicate if \textbf{Ours} scored higher than \textbf{Informative}, where $*$ denotes statistical significance.}
	\label{table:likert}
	\centering
		\begin{tabular}{lcccc}
			\hline Questionnaire Item & Reliability & $F(1,78)$ & p-value \bigstrut \\ \hline 
            
            -- I found it \textbf{\emph{easy}} to answer the robot's questions. & \multirow{2}{*}{$.85$} & \multirow{2}{*}{$11.12$} & \multirow{2}{*}{$p<.01^*$} \\  -- The robot's questions were hard to answer. \bigstrut[b] \\ \hline  
            
            -- The robot learned from my answers and \textbf{\textit{improved}}. & \multirow{2}{*}{$.25$} & \multirow{2}{*}{$-$} & \multirow{2}{*}{$-$} \\ -- The robot kept asking things I thought it already knew. \bigstrut[b] \\ \hline  

            \multirow{2}{22em}{-- By the end of the task the robot's questions \textbf{\textit{revealed}} what the robot had learned.} & \multirow{4}{*}{$.78$} & \multirow{4}{*}{$10.55$} & \multirow{4}{*}{$p<.01^*$} \\ \\ \multirow{2}{22em}{-- At the end of the task it was still unclear what the robot had and had not learned.} \\ \bigstrut[b] \\ \hline  
            
            \multirow{2}{22em}{-- The robot \textbf{\textit{focused}} on specific parts of the task when asking questions.} & \multirow{3}{*}{$.93$} & \multirow{3}{*}{$27.81$} & \multirow{3}{*}{$p<.01^*$} \\ \\ -- The robot always seemed unsure about the entire task. \bigstrut[b] \\ \hline  
            
            -- The robot's \textbf{\textit{learned behavior}} matched my preferences. & \multirow{2}{*}{$.92$} & \multirow{2}{*}{$1.44$} & \multirow{2}{*}{$p=.233$} \\ -- The robot did not learn my preferred trajectory. \bigstrut[b] \\ \hline

		\end{tabular}

\vspace{-0.5em}

\end{table*}

\p{Results -- Objective} Convergence (see \fig{user_study}) highlights one key difference between the questions generated by \textbf{Ours} and \textbf{Informative}. With \textbf{Informative}, the robot selects queries that are often far from what the robot has learned, potentially \textit{misleading} the human into thinking the robot is confused or unsure. By contrast, with \textbf{Ours} the robot asks questions that gradually hone-in on what the robot has learned, demonstrating this learning to the human. We emphasize that these experimental results are consistent with Proposition~1, since here the questions generated by \textbf{Ours} become more revealing over time (i.e., convergence decreases, indicating more revealing questions). Across all three tasks our objective results support \textbf{H2}, and suggest that robots which account for the bi-directional transfer of information purposely reveal their learned behavior to the human.

\subsection{Follow-up: Bringing the Human into the Learning Process}

The results of the first user study indicate that our approach improves the human's perception of the robot. But what are the \textit{practical benefits} of keeping the human up-to-date with the robot's learning? In this second user study we test two potential benefits: i) determining which features the robot is still unsure about, and ii) determining when the robot learner is ready to be deployed. Overall, the purpose of this second user study is to see whether bringing the human into the learning loop can improve the interactive learning process.

\p{Participants and Procedure} We recruited 10 new community members (6 female, ages $25 \pm 3.5 $ years) to participate in this follow-up study. As before, participants provided informed written consent following Virginia Tech IRB \#20-755. These participants completed the \textit{Sort} and \textit{Stack} tasks from \fig{user_study} with some slight modifications (described below). For both tasks the robot was initially unsure about the human's true preferences $\theta^*$. 

In the modified \textit{Sort} task the robot was trying to learn three preferences: which \textit{cup} to pick up, which \textit{shelf} to sort this cup onto, and whether to place the cup in the front or back of the shelf (\textit{distance}). We artificially limited the preferences that the robot could learn from the human's answers: no matter what the human chose, the robot never learned which \textit{cup} to pick up. After a total of twelve robot questions we asked participants to specify which preferences they thought the robot had learned and which preferences the robot was still unsure about.

By contrast, in the modified \textit{Stack} task we did not impose any limits on the robot's learning. Here participants taught the robot until they thought it had learned their preferences and was ready to be autonomously deployed. Each time the robot asked a question participants were given the option to stop teaching and deploy the robot. We recorded the round where participants chose to discontinue teaching (up to a maximum of twelve questions).

Participants completed the modified tasks once with the \textbf{Informative} robot and once with \textbf{Our} approach. As before, we counterbalanced the order of presentation for this within-subjects design.

\begin{figure}[t]
	\begin{center}
		\includegraphics[width=0.8\columnwidth]{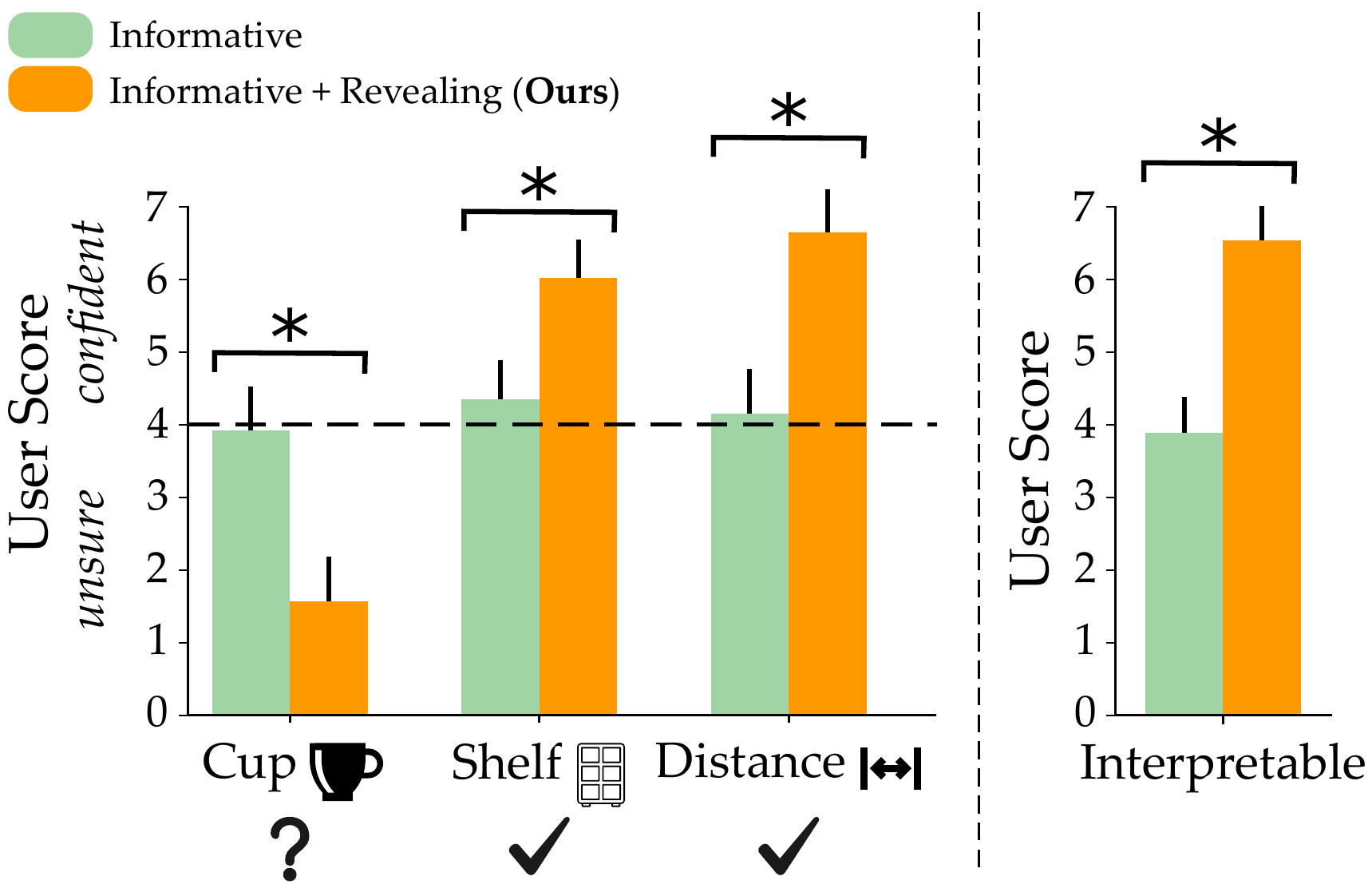}
		\caption{User perceptions of what the robot did and did not learn during the modified \textit{Sort} task. In this task the robot was only able to learn \textit{shelf} and \textit{distance} preferences, and we artificially prevented the robot from learning which \textit{cup} to sort. (Left) High scores indicate the human thought the robot learned a preference, while low scores indicate the human thought the robot was confused. With the state-of-the-art \textbf{Informative} baseline participants were on the fence for every preference, even preferences the robot actually learned. (Right) We also asked participants whether the overall behavior of each robot was interpretable. Higher ratings indicate agreement, and $*$ denotes statistical significance ($p < .001$).}
		\label{fig:subjective_s2}
	\end{center}

\end{figure}

\p{Dependent Measures -- Revealing Learning} Our analysis for the \textit{Sort} task was similar to the online user studies from Section~\ref{sec:amt}. After answering all twelve questions participants indicated what they thought the robot had learned about each preference on a $7$-point scale. Here a score of $1$ denotes that the human believed the robot never learned that preference, while a score of $7$ denotes that the human thought the robot fully understood what the they wanted. We also asked participants whether the robot's queries were \textit{Interpretable} throughout the entire interaction.

\p{Dependent Measures -- Ready to Deploy} During the \textit{Stack} task we recorded how many questions the human answered before deciding that their robot was ready to be deployed (\textit{Deploy}). Of course, the human might think the robot learned their preferences before it actually did --- accordingly, we also recorded the first question where the robot's regret was below a threshold. Recall from Section~\ref{sec:sim} that regret is the difference in reward between the human's preferred behavior and the robot's learned behavior. By recording regret we thus identified a lower bound on when the robot was actually ready to be deployed.

\p{Hypotheses} We had the following two hypotheses:
\begin{displayquote}
    \textbf{H3.} \emph{Humans will be misled by robots that only optimize for informative questions, but will recognize what the robot has and has not learned under our approach.} 
\end{displayquote}
\begin{displayquote}
    \textbf{H4.} \emph{Our approach will cause participants to deploy the robot soon after it has learned their preferences without answering additional and unnecessary questions.} 
\end{displayquote}

\p{Results -- Revealing Learning} Our results on the modified \textit{Sort} task are shown in \fig{subjective_s2}. Recall that here we artificially prevented the robot from learning the \textit{cup} feature. When participants interacted with the \textbf{Informative} robot they did not realise this. Indeed, in the \textbf{Informative} condition participants were misled about every preference, and were unsure if the robot even understood their preferred \textit{shelf} or \textit{distance}. But with \textbf{Our} approach participants got it right: they correctly recognizes what the robot did not learn (\textit{cup}) and what the robot was confident about (\textit{shelf} and \textit{distance}). The difference between these interpretations was significant across three features ($p<.001$). Overall, participants perceived the questions asked by \textbf{Our} robot as more interpretable than the questions asked by the state-of-the-art \textbf{Informative} robot ($p<.001$).

\begin{figure}[t]
	\begin{center}
		\includegraphics[width=1\columnwidth]{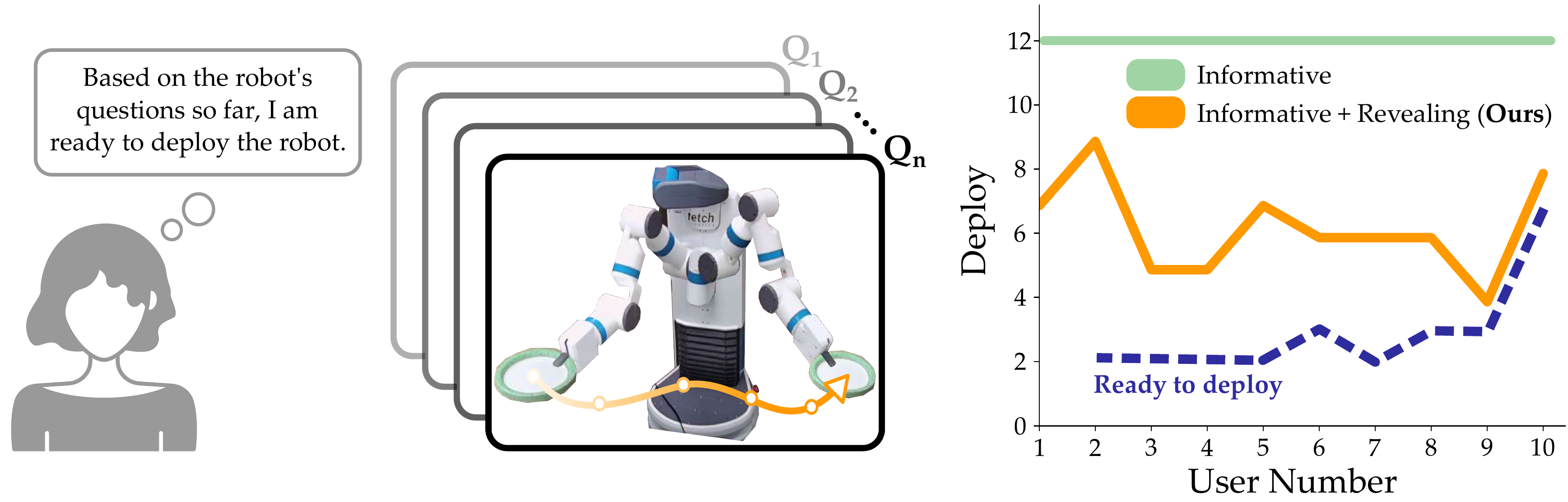}
		\caption{Participants determining when the robot learner is ready for autonomous deployment. (Left) Participants could choose to deploy the robot after any question, up to a maximum of twelve questions. (Right) We measured the robot's regret to determine when it was actually ready to be deployed. The dashed line shows when this regret was less than a threshold of $0.2$, indicating that the robot understood the human's preferences. Ideally participants will chose to deploy the robot immediately after this dashed line. With \textbf{Informative} participants were never confident in the robot and did not chose to deploy it (answering many unnecessary questions). By contrast, the questions asked by \textbf{Ours} revealed that the robot had learned the human's desired behavior, and so participants chose to deploy this robot after around six questions. User $1$ changed their preferences multiple times, making it unclear when this robot was ready to be deployed.}
		\label{fig:questions_s2}
	\end{center}
	
	\vspace{-1em}

\end{figure}

\p{Results -- Ready to Deploy} Our results on the modified \textit{Stack} task are visualized in \fig{questions_s2}. Here we recorded how many questions each user thought they needed to teach the robot their desired behavior. The dashed line shows when the robot was actually ready to be deployed: the robot typically understood how to carry and stack the dishes after between two and four questions. However, the robot's learning was not transparent in the \textbf{Informative} condition. Here participants were never confident that the robot understood their preference, and so they answered all twelve questions without choosing to deploy the robot. By contrast, the questions that \textbf{Our} robot asked helped participants recognize that it was ready to be deployed. Participants consistently deployed \textbf{Our} robot after it had successfully learned their preferences. These results support \textbf{H4} and indicate that --- by actively bringing the human into the learning loop --- robots using our approach encourage humans to trust and deploy learning agents.

%% file: conclusion.tex
\section{Conclusion}

We explored settings where a robot is asking questions to learn the human's reward function. Our key insight was that robot questions implicitly convey information to the human, revealing what the robot does and does not know. Prior work on active robot learning studies a one-way transfer of information: from the human to the robot. In this paper we considered both human and robot perspectives, and developed a method for asking questions that closes the loop and conveys what the robot has learned back to the human teacher.

We first formalized the bi-directional transfer of information that occurs when robots ask questions. Here we introduced a cognitive human model that relates the questions the robot asks to the information those questions reveal. We tested this revealing model in online user studies, where $50$ participants were able to correctly identify what the robot did and did not know based only on the questions the robot asked.

Next, we formulated a continuous trade-off between informative and revealing questions. We found that when robots only optimize for gathering information users perceive their questions as random and misleading. Conversely, when robots only optimize for revealing information, they learn just as slowly as a robot asking random questions. Under our proposed approach robots optimize for both human and robot perspectives: these robots actively select questions to efficiently learn in an interpretable manner.

We conducted simulations and user studies to compare our approach to a state-of-the-art alternative. Robots leveraging our algorithm learned as quickly as robots that optimize for information gain, but with the added benefit of revealing that learning to the human. In practice, this enabled participants to i) identify which parts of the task the robot was still unsure about, and ii) determine when their robot had learned their preferences and was ready to be deployed.

%% file: limitations.tex
\section{Limitations and Future Work} 

Our work is a step towards interpretable and transparent active learning robots. However, we recognize that there are many modalities that robots can leverage to communicate their learning to the human: e.g., natural language, haptic interfaces, or augmented reality. In this paper we specifically focused on using robot motion, but future work should consider multiple, complementary types of feedback. Depending on the situation one or more of these modalities may be preferable.

During our user studies participants commented that they were unsure about which part of the robot's question to focus on. Returning to our motivating example, imagine that the robot is uncertain about both carrying height and stacking the dishes. The robot shows two trajectories: one gets the height wrong but correctly stacks the dishes, and the other gets the height right but fails to stack correctly. Which option should the user choose? We expected users to select the preference that was more important to them (either height or stacking), but some participants voiced confusion in these scenarios. This indicated to us that the robot should focus on one preference at a time --- however, in our follow-up online survey, $50$ participants perceived the all-at-one and one-at-a-time robots as similarly revealing. Moving forward, our working hypothesis is that robot questions can vary multiple parts of the task when \textit{revealing} information, but should focus on a single preference when \textit{gathering} information. We plan to explore this point in future work.